\documentclass[manuscript]{aastex}

\usepackage{multirow}
\usepackage{multicol}

\shorttitle{THE TEMPORAL AND SPECTRAL CHARACTERISTICS OF ``FAST RISE
AND EXPONENTIAL DECAY'' GAMMA-RAY BURST PULSES} \shortauthors{Peng
et al.}

\begin{document}


\title{THE TEMPORAL AND SPECTRAL CHARACTERISTICS OF ``FAST RISE AND EXPONENTIAL DECAY'' GAMMA-RAY BURST PULSES}

\author{Z. Y. Peng \altaffilmark{1}, Y. Yin\altaffilmark{2}, X. W. Bi\altaffilmark{3}, X. H. Zhao\altaffilmark{4},
L. M. Fang\altaffilmark{5}, Y. Y. Bao\altaffilmark{6}, L. Ma
\altaffilmark{1,{\ast}}}

\altaffiltext{1}{Department of Physics, Yunnan Normal University,
Kunming 650092, China; pzy@ynao.ac.cn}

\altaffiltext{${\ast}$}{Corresponding author, astromali@126.com}

\altaffiltext{2}{Department of Physics, Liupanshui Normal College,
Liupanshui 553004, China}

\altaffiltext{3}{Department of Physics, Honghe College, Mengzi,
661100, China}

\altaffiltext{4}{National Astronomical Observatories/Yunnan
Observatory, Chinese Academy of Sciences, P. O. Box 110, Kunming
650011, China}

\altaffiltext{5}{Department of Physics, Guangdong Institute of
Education, Guangzhou 510303, China}

\altaffiltext{6}{Department of Physics, Yuxi Normal College, Yuxi
653100, China}


\begin{abstract}
In this paper we have analyzed the temporal and spectral behavior of
52 Fast Rise and Exponential Decay (FRED) pulses in 48 long-duration
gamma-ray bursts (GRBs) observed by the CGRO/BATSE, using a pulse
model with two shape parameters and the Band model with three shape
parameters, respectively. It is found that these FRED pulses are
distinguished both temporally and spectrally from those in long-lag
pulses. Different from these long-lag pulses only one parameter pair
indicates an evident correlation among the five parameters, which
suggests that at least $\sim$4 parameters are needed to model burst
temporal and spectral behavior. In addition, our studies reveal that
these FRED pulses have correlated properties: (i) long-duration
pulses have harder spectra and are less luminous than short-duration
pulses; (ii) the more asymmetric the pulses are the steeper the
evolutionary curves of the peak energy ($E_{p}$) in the $\nu
f_{\nu}$ spectrum within pulse decay phase are. Our statistical
results give some constrains on the current GRB models.



\end{abstract}

\section{INTRODUCTION}
The temporal profiles of Gamma-ray burst (GRB) are very diverse in
morphology but the spectra could be fitted with a single simple Band
model (Band et al. 1993). The spectral parameters (the power-law
indices and the peak energy in the $\nu f_{\nu}$ spectrum) are then
used to infer the GRB emission and particle acceleration mechanisms.
However, the signatures of the gamma-ray epoch of the burst are
hidden in the time evolution of the light curve and in its spectral
behavior. The individual emission episodes (pulses) that complex
light curves are believed to consist of reflect the behavior of
central engine. Due to the overlapping of pulses in most bursts,
especially that bright ones, only a small fraction of all bursts
consist of long, smooth and well-shaped pulse, often with a fast
rise and a exponential decay (FRED), while others exhibit very
complex and jagged light curves. Therefore accurate study of
individual pulse behavior is often difficult. However, some dimmer
bursts with lower signal-to-noise ratios usually have simpler
temporal structure may be easy to model (Norris et al. 2005,
hereafter Paper I). The investigations of these pulses are useful,
which might lead to a deeper understanding of the creation of the
gamma-rays by giving clues to and constraining physical models.

Many authors have studied the temporal and spectral properties of
long-duration ($T_{90} > 2$ s) GRB pulses, a number of
characteristics of these pulses are revealed. The impressive results
include: e.g. (1) the temporal asymmetry of pulses in GRBs, that is
longer decay than rise rates, (2) hard-to-soft spectral evolution,
and (3) energy dependence of the pulse duration, broadening at lower
energies (e.g., Norris et al. 1996; Ryde 2005; Golenetskii et al.
1983; Borgonovo \& Ryde 2001; Kouveliotou et al. 1993).

However, most of these studies focus on the pulses in bright bursts.
Stern et al. (1999) investigated a complexity-brightness correlation
in GRB and found that the average profiles of dim bursts were less
complex than that of bright bursts. Based on this Paper I analyzed
the temporal and spectral behavior of some long-lag bursts, which
tend to be dim but also to have relatively simple temporal
structures. They found that pulses in long-lag bursts are
distinguished both temporally and spectrally from those in bright
bursts: (1) the pulses in long-lag bursts are few in number, (2) the
durations are $\sim$ 100 times wider (tens of seconds) than those of
bright bursts, (3) the peak energy $E_{p}$ in $\nu f(\nu)$ is lower,
and (4) the long-lag bursts have harder low-energy spectra and
softer high-energy spectra.

Kocevski et al. (2003) analyzed the time profiles of 76 FRED pulses
with the peak flux greater than that long-lag pulses. They only
considered the temporal behavior of these pulses and did not analyze
the spectral properties. Employing this sample Peng et al. (2009a)
(hereafter Paper II) studied the spectral behavior of these FRED
pulses that are bright enough to perform spectral analysis. They
focused their attentions on the evolutionary slope of peak energy
$E_{p}$ within the pulse decay phase and found that the slope is
correlated with several spectral parameters.

In the present work, we would like to employ the sample presented by
Paper II to investigate the temporal and spectral properties of
these FRED pulses. If these bursts are distinguished from long-lag
bursts temporally and spectrally is our another motivation. In
Section 2, we present the sample description. The temporal and
spectral profile analysis are given in Section 3. Discussion and
conclusions are presented in the last section.

\section{SAMPLE DESCRIPTION}

Paper II used two samples provided by Kocevski et al. (2003) and
Norris et al. (1999) to investigate the evolutionary slope of
$E_{p}$ in FRED pulses. The main selected criterions of the two
samples of Paper II are: (1) the data are provided by the BATSE
instruments on board the CGRO spacecraft and the duration is greater
than 2 s ($T_{90} > 2$ s); (2) exhibited clean, single-peaked events
or, in the case of multi-peaked bursts, pulses that were well
distinguished and separable from each other; (3) the peak flux is
greater than 1.8 photon $cm^{-2} s^{-1}$ on a 256 ms timescale. The
time-resolved and time-integrated spectra of the two samples were
fitted with the Band and Compton model, respectively. Based on these
fitting parameters Paper II studied the evolutionary slope of
$E_{p}$ as well as the correlations between the slope and the
spectral parameters. Their analysis showed the two samples share
approximately the same statistical properties, which can be found
from Figure 1 to Figure 9 in Paper II (for more details of the
samples and the spectral modeling, one can refer to Kocevski et al.
2003, Norris et al. 1999, and Paper II). Therefore, we only select,
in this paper, the sample fitted by Band model to investigate the
temporal and spectral characteristics of FRED pulses, which includes
56 single pulses.

\section{TEMPORAL PROFILE AND TIME-INTEGRATED SPECTRA  ANALYSIS}

\subsection{Temporal Profile Analysis}

Once these pulses have been selected we use the pulse model of Paper
I to fit them. The pulse model can be rewritten as follows:
\begin{equation}
I(t)= A \lambda \exp [-\tau_{1} /(t- t_{s})-(t - t_{s})/\tau_{2} ],
\end{equation}
where $t$ is time since trigger, $A$ is the pulse amplitude, $t_{s}$
is the pulse start time, $\tau_{1}$ and $\tau_{2}$ are
characteristics of the pulse rise and pulse decay, and $\lambda
=\exp[2(\tau_{1}/\tau_{2})]^{1/2}$.

Similar to Peng et al. (2006, 2009b) and Hakkila et al. (2008) we
also use nonlinear least squares routine MPFIT to fit these pulses.
It is based on the well-known and tested MINPACK-1 FORTRAN package
of routines available at www.netlib.org. Moreover, MPFIT functions
may permit you to fix any function parameters, as well as to set
simple upper and lower parameter bounds. To obtain an intuitive view
of the result of the fit, we develop and apply an interactive IDL
routine for fitting pulses in bursts, which allows the user to set
and adjust the initial pulse parameters manually before allowing the
fitting routine to converge on the best-fitting model via the
reduced $\chi^{2}$ minimization. The background-subtracted light
curves combined four channels are fitted with the pulse model. The
fits are examined many times to ensure that they are indeed the best
ones. The fitting $\chi^{2}$ per degree of freedom larger than 2.5
are rejected. In the end there are 52 pulses are included in our
sample.

We demonstrate two fit results with the largest values of $\chi^{2}$
(GRB 980301 (BATSE trigger 6621)) and with smallest value (GRB
931128 (BATSE trigger 2665)) in Figure 1. The distributions of
$\chi^{2}$ per degree of freedom for our sample are displayed in
Figure 2. The narrow distribution of the $\chi^{2}$ values indicates
that two-exponential model is sufficient to model the pulse light
curves.

\begin{figure}
\centering \resizebox{3in}{!}{\includegraphics{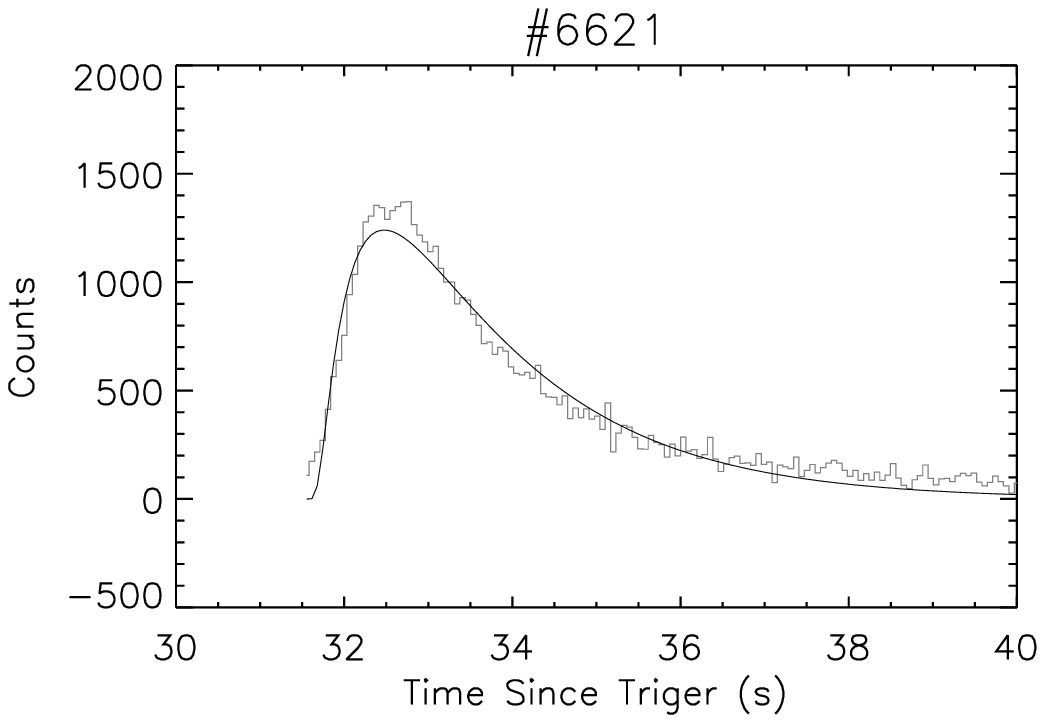}}
\resizebox{3in}{!}{\includegraphics{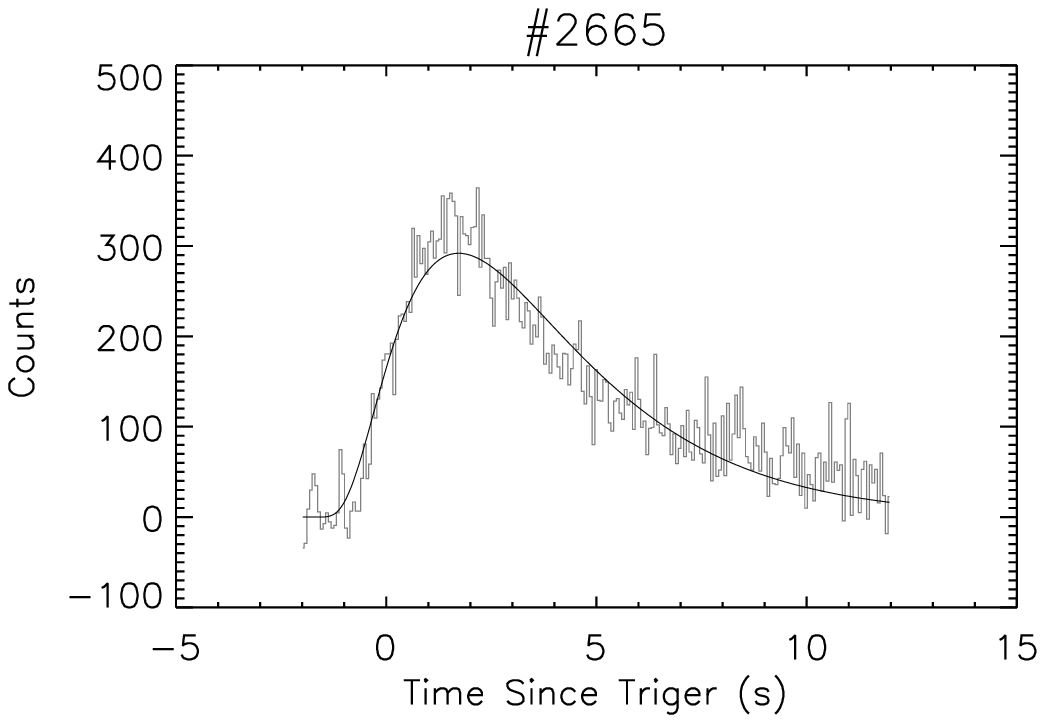}} \caption{The plots of
the fitting result of two pulse with the largest value of $\chi^{2}$
(left panel) and with the smallest value of $\chi^{2}$ (right panel)
 in our sample.}
 \label{}
\end{figure}

\begin{figure}
\centering \resizebox{3in}{!}{\includegraphics{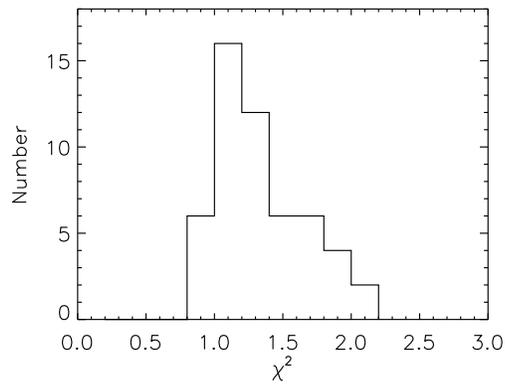}}
\caption{Histograms for the distribution of $\chi^{2}$ in our
sample.} \label{}
\end{figure}

According to the fitted parameters we can obtain the two shape
parameters, width $w$ and asymmetry $k$. Following Paper I we find
the pulse width measured between the two $1/e$ intensity points, $w
= \bigtriangleup \tau _{1/e} =\tau _{2}(1+2\ln \lambda)^{1/2} $. The
form of the pulse asymmetry $k=\tau _{2}/w$. Quilligan et al. (2002)
found that the full-widths at half-maximum (FWHM) of GRB pulse is
log-normal distribution. It is found the distribution of $w$ is also
log-normal (see Figure 3) but the distribution of $k$ is normal
(Figure 3). The parameters of the best log-normal and normal fits
are given in Table 1.

In addition, we find no significant correlation between the width
and asymmetry, which is consistent with the result of long-lag
pulses. The widths of these FRED pulses are distinguished from the
long-lag pulses since the analysis of long-lag pulse performed by
Paper I showed the average width is larger than 10 s. Whereas the
difference of the mean value of the asymmetry between the FRED
pulses and the long-lag pulses is not evident (see, Table 1).
Actually considering the standard deviations ($\sigma$) and the
number of pulses in each sample (\#52 FREDs and \#35 long-lag from
Paper I) the mean k's are within 1 standard error of the sample
mean. In other words, they are equal within uncertainties.


\begin{figure}
\centering \resizebox{3in}{!}{\includegraphics{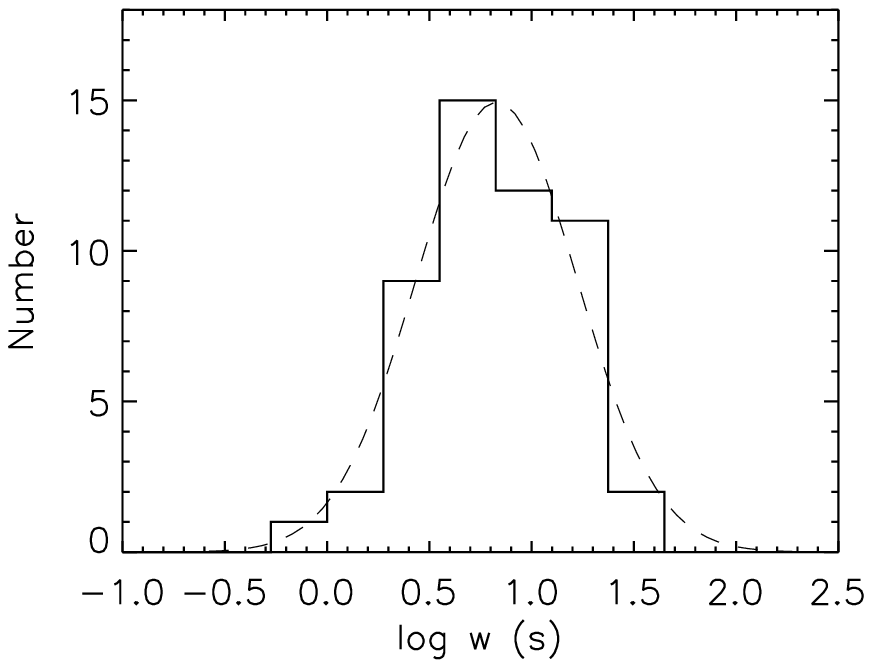}}
\resizebox{3in}{!}{\includegraphics{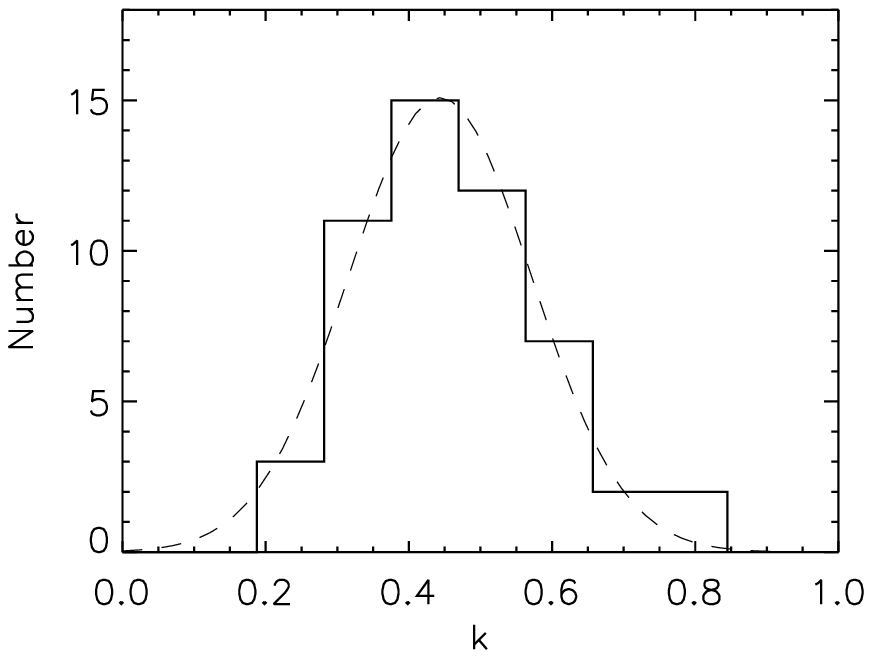}} \caption{Distributions
of the pulse width, $w$ (left panel) and pulse asymmetry, $k$ (right
panel) in our sample, where the curves represent the Gaussian fit to
the two distributions.} \label{}
\end{figure}

\begin{deluxetable}{ccccccc}
\tabletypesize{\scriptsize}  \tablecaption{A comparison of temporal
and spectral parameters for the FRED and long-lag pulses.
\label{tbl-1}}\tablewidth{0pt} \tablehead{\multirow{2}{*}{} &
\multicolumn{2}{c}{FRED pulse} & &\multicolumn{2}{c}{Long-lag pulse$^{a}$} \\
\cline{2-3} \cline{5-6}  \colhead{property} & \colhead{$\mu$} &
\colhead{$median$}& &\colhead{$\mu$} & \colhead{$median$}}
\startdata
$w$ (s) & 6.74 $\pm$ 2.47 & 6.45 & & 15.18$\pm$ 13.75 & 12.96\\
$k$ & 0.44 $\pm$ 0.13 & 0.46 &  & 0.40 $\pm$ 0.19 & 0.41 \\
$E_{p}$ (keV) & 158.49 $\pm$ 79.67 & 161.40 & &109.89 $\pm$64.50 & 110.75 \\
$\alpha$ & -0.89 $\pm$ 0.52 & -0.87 & & -0.46 $\pm$ 0.65 & -0.47\\
$\beta$ & -2.60 $\pm$ 0.37 & -2.63 && -2.74 $\pm$0.25 & -2.80\\
\enddata
\tablecomments{$^a$ Reference for the long-lag pulse data: Norris et
al. (2005), i.e., Paper I.}
\end{deluxetable}

\subsection{Spectral Profile Analysis}

Paper II described in detail the spectral modeling for the single
pulses and analyzed the evolutionary slope, $S$, during the decay
phase of the FRED pulses. In addition, they examined the relations
between spectral parameters and $S$. In this section, let us first
check the distributions of time-integrated spectral parameters for
our sample. Figure 4 indicates the distributions of the spectral
parameters, $E_{p}$, $\alpha$, $\beta$. The corresponding mean value
and standard deviation are listed in Table 1. Note that the spectral
parameters come from spectra integrated over a FRED pulse rather
than an entire burst.

\begin{figure}
\centering \resizebox{3in}{!}{\includegraphics{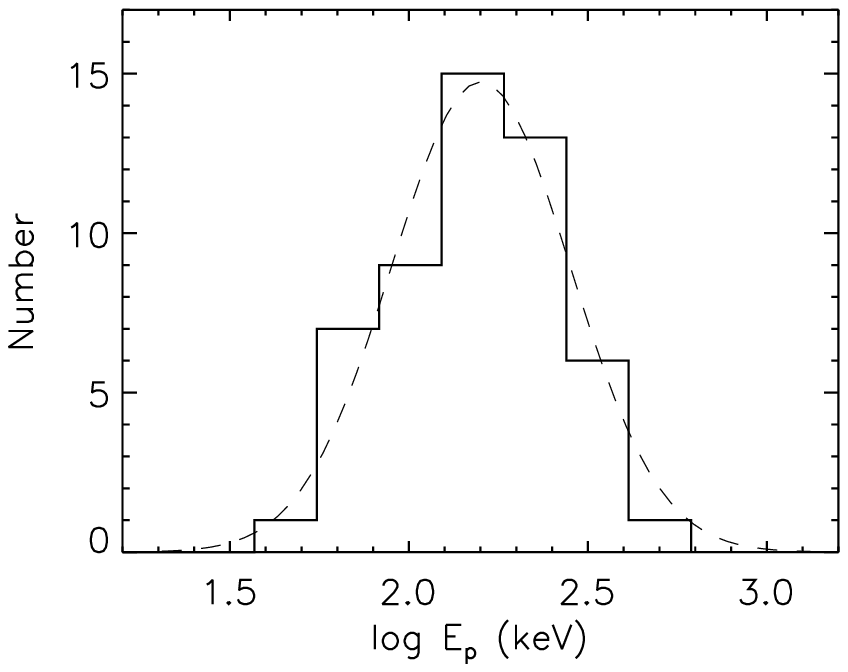}}
\resizebox{3in}{!}{\includegraphics{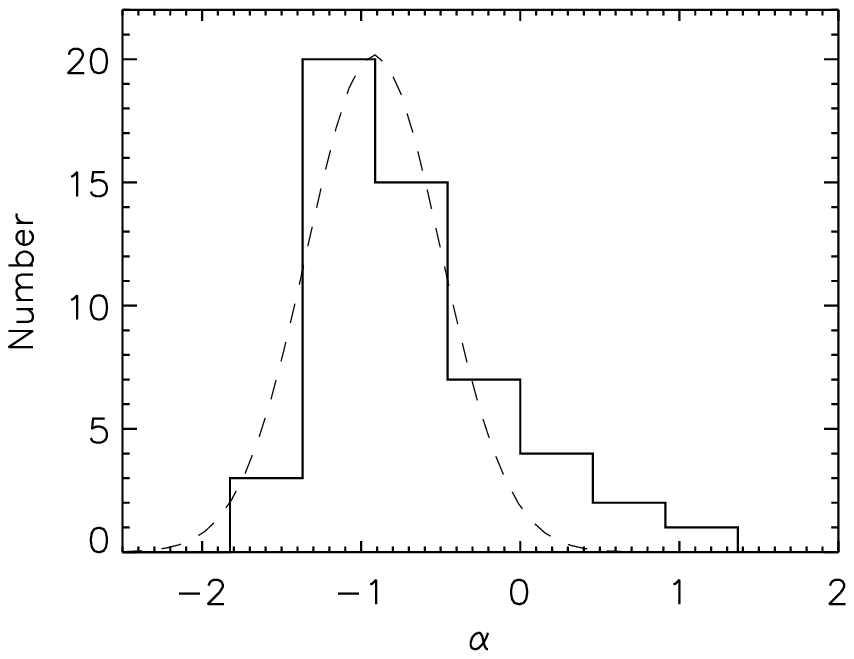}}
\resizebox{3in}{!}{\includegraphics{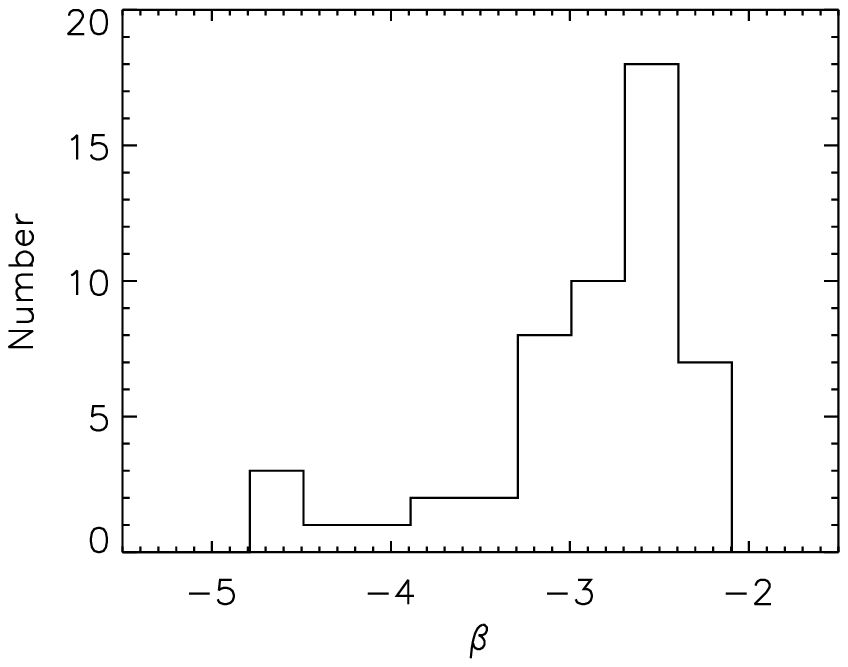}}
\caption{Distributions of the pulse peak energy, $E_{p}$, low-energy
index, $\alpha$, and high-energy index, $\beta$ in our sample, where
the curves represent the Gaussian fit to the two distributions.}
\label{}
\end{figure}

Previous study showed the distribution of $E_{p}$ integrated over a
burst is best described by a log-normal distribution (Quilligan et
al. 2002). We find $E_{p}$ distribution for the single pulses is
also log-normal. Compared with these long-lag pulse the $E_{p}$ of
FRED pulse is much greater (see Table 1). While for the low-energy
index the distribution is approximately normal and the spectra of
most FRED pulses are much steeper than those of the long-lag pulses.
Whereas for the high-energy index the spectra of most of FRED pulses
are a little flatter than those of the long-lag pulses (see Table
1).

\subsection{The Relation Between The Temporal And Spectral Parameters}
Paper I have examined the relations between the temporal and
spectral parameters and showed: (1) no clear correlation between the
low-energy index $\alpha$ and the width of pulse $w$ is indicated;
(2) there is a suggestion that $\alpha$ is correlated with pulse
asymmetry; (3) the $E_{p}$ appears to be uncorrelated with any
temporal parameters; (4) neither of two temporal parameters is
correlated with high-energy index $\beta$. We re-examine the
relations and find that the results are also established for our
sample except that two differences from that of long-lag pulse
(Paper I). The first difference is that the relation between
$\alpha$ and $w$. A correlated relation between them is suggested
for our sample (see, Figure 5 and Table 2). The second difference
that there seems no correlation between $\alpha$ and $k$ (see Figure
5 and Table 2). The other parameter pairs, $E_{p}$ versus $\alpha$,
$E_{p}$ versus $\beta$, $E_{p}$ versus $w$, $E_{p}$ versus $k$,
$\alpha$ versus $\beta$, $w$ versus $\beta$, and $k$ versus $\beta$,
are not correlated with each other.

\begin{figure}
\centering \resizebox{3in}{!}{\includegraphics{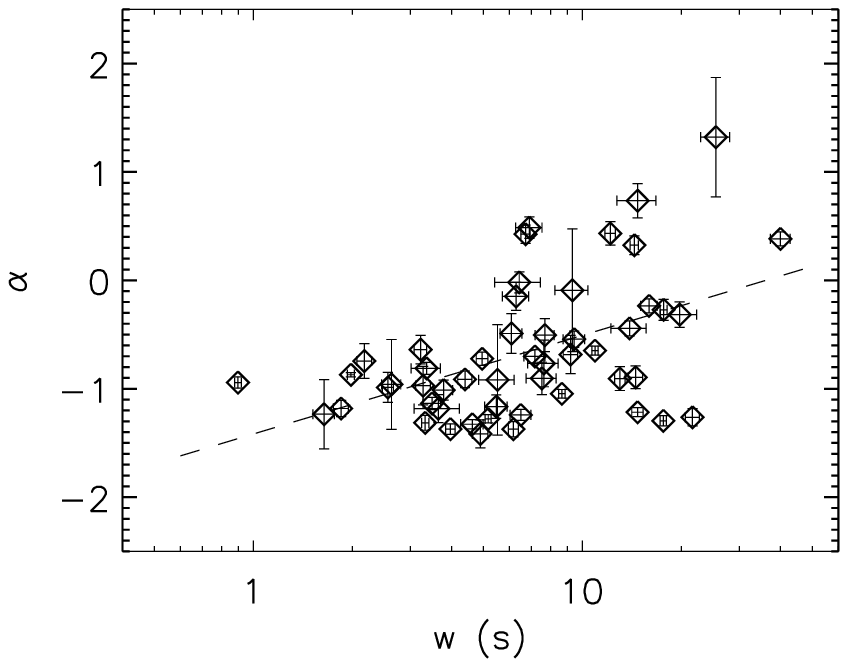}}
\resizebox{3in}{!}{\includegraphics{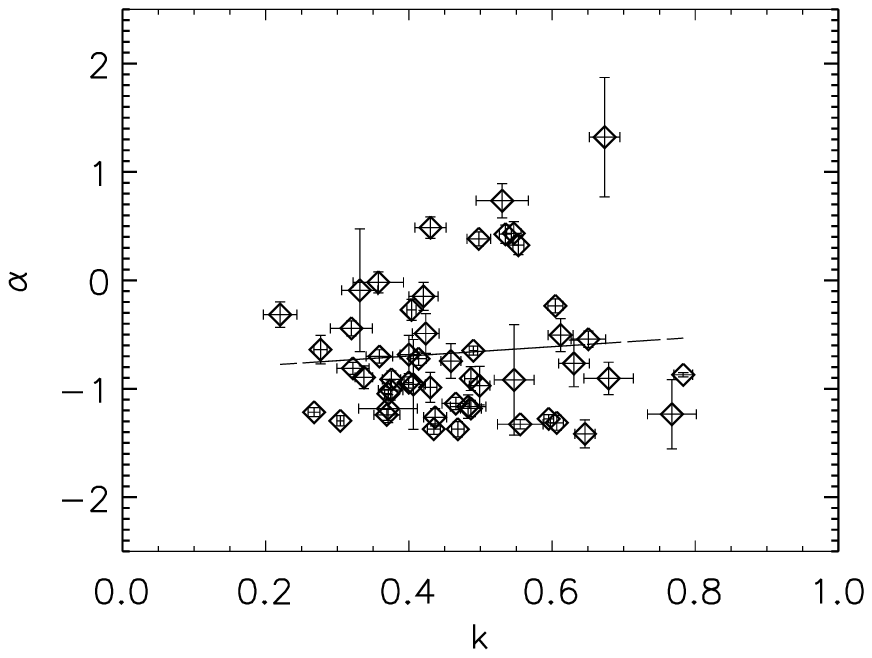}} \caption{Spectral
shape parameter $\alpha$ vs. pulse width $w$ (left panel) as well as
$\alpha$ vs. pulse asymmetry $k$ (right panel), where the long
dashed lines represent the best fitting lines.} \label{}
\end{figure}

\begin{deluxetable}{ccc}
\tabletypesize{\scriptsize}  \tablecaption{Correlations of the six
parameter pairs. \label{tbl-1}}\tablewidth{0pt} \tablehead{
\colhead{Parameter Pair} & \colhead{$R_{S}$} & \colhead{$P_{S}$} }
\startdata
$w-\alpha$ & 0.49 & 0.0006 \\
$k-\alpha$ & -0.0018 & 0.99 \\
$w-S$  & -0.28 & 0.04  \\
$k-S$ & 0.49 & $ < 10^{-4}$   \\
$w-F$ & -0.53 & $ < 10^{-4}$  \\
$k-F$ & -0.17 & 0.22 \\
\enddata
\end{deluxetable}


Paper II have studied the relations between the decay slope of
pulse, $S$, and the spectral parameters. In this work we mainly
check the relations between $S$ and two temporal parameters as well
as photon flux and temporal parameters.

\begin{figure}
\centering \resizebox{3in}{!}{\includegraphics{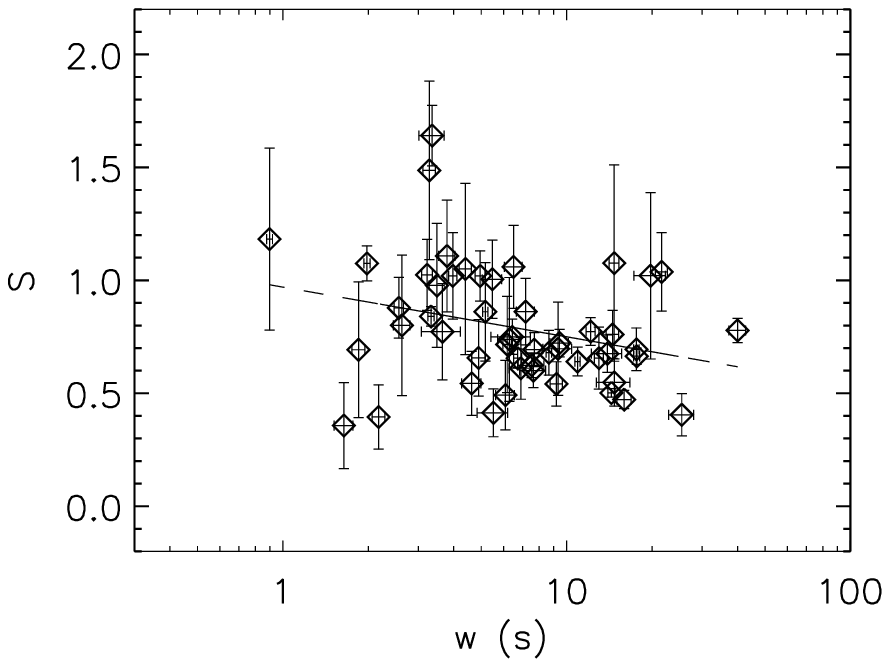}}
\resizebox{3in}{!}{\includegraphics{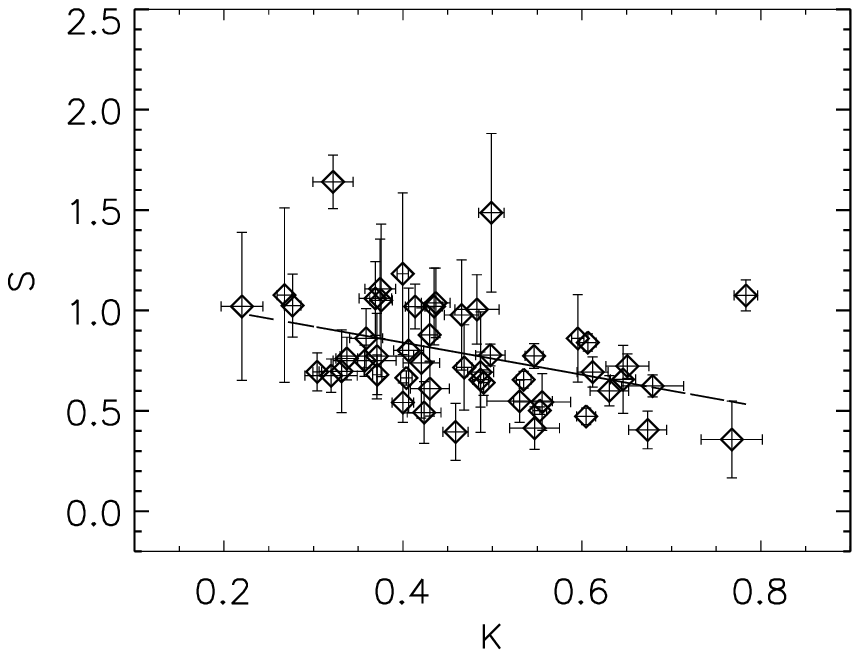}} \caption{The $E_{p}$
evolutionary slope during the pulse decay phase, $S$, vs. pulse
width $w$ (left panel) and pulse asymmetry $k$ (right panel), where
the long dashed lines represent the best fitting lines.} \label{}
\end{figure}

Figure 6 (left panel) shows the relation between the $S$ and $w$. A
anti-correlation between them is identified for our samples (also
see Table 2). In addition, a correlation between the $S$ and pulse
asymmetry $k$ is suggested in Figure 6 (right panel) and Table 2.
Figure 7 shows a similar picture for the photon flux versus two
temporal parameters. A clear correlated relation between photon flux
and $w$ is identified, but there seems no evident correlation
between the photon flux and $k$ (also see Table 2).

\begin{figure}
\centering \resizebox{3.2in}{!}{\includegraphics{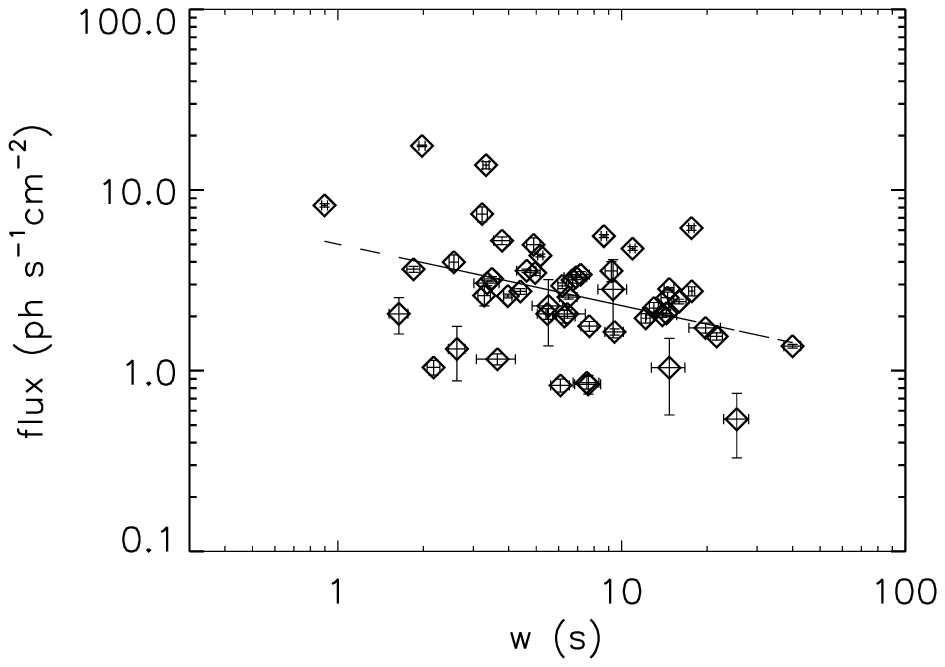}}
\resizebox{3.2in}{!}{\includegraphics{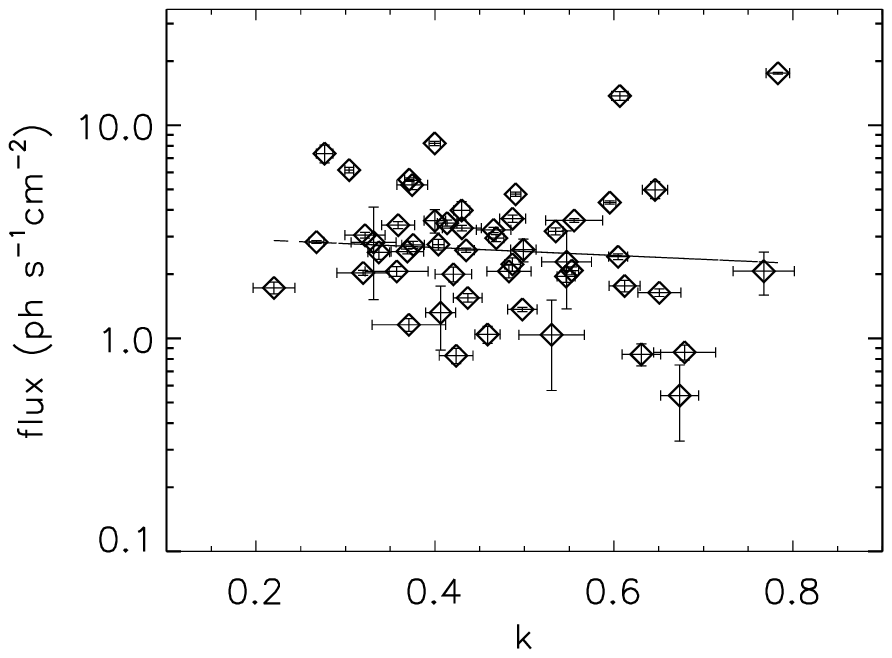}} \caption{The photon
flux vs. pulse width, $w$, (top panel) as well as photon flux vs.
pulse asymmetry, $k$, for our sample, where the long dashed lines
represent the best fitting lines.} \label{}
\end{figure}

Based on the above analysis the five fundamental temporal and
spectral shape parameters $w$, $k$, $\alpha$, $\beta$, and $E_{p}$
do not show compelling evidence for any pairwise correlation, except
for a correlation between $w$ and $\alpha$. This implies that at
least four independent physical parameters are required to determine
pulse behavior in the energy band $\sim $25 - 2000 $keV$, which is
different from that of long-lag pulse investigated by Paper I.

\section{Discussion and Conclusions}
By studying the temporal and spectral characteristics of FRED GRB
pulses we first show that the FRED pulse is distinct from that
long-lag pulse temporally and spectrally: (1) the average width of
FRED pulses (8.75 s) is below the corresponding values of long-lag
(the average width is above 10 s, Paper I); (2) the average peak
energy, $E_{p}$, in the $\nu f _{\nu}$ is 158 keV which is also
greater than the long-lag pulse (110 keV); (3) the low-energy
indices obtained from FRED pulses are softer than that of long-lag
pulses; (4) while the high-energy indices of FRED pulses are
slightly harder than that of long-lag pulses. Therefore, these
long-lag, wide-pulse GRBs and these general sample of GRB pulses may
represent different sub-class with generally different physical
properties. But the difference of the pulse asymmetry between the
FRED and long-lag pulses is not significant.

Analysis of relations of temporal and spectral shape parameters
suggests that they have no visible correlation except that the pulse
width is correlated with lower-energy index $\alpha$, which
indicates that at least four parameters are needed to model burst
temporal and spectral behavior. The inconsistency of correlations
between $w$ and $\alpha$ as well as pulse asymmetry $k$ and $\alpha$
with that of the long-lag burst studied by Paper I may be caused by
the sample size. Our sample consisting of 52 pulses is a factor of 2
larger than that of long-lag bursts. Ryde et al. (2005) and Ryde
(2005) showed a similar relation that hard spectra (with large
spectral power-law indices $\alpha$) give the largest lags.
Moreover, Paper I pointed out that pulse width is strongly
correlated with spectral lag and these two parameters may be viewed
as mutual surrogates. If this is the case we tend to believe that
there are indeed correlation between $w$ and $\alpha$. Our analysis
confirms that $k$ is not correlated with $\alpha$. Another
possibility is that the characteristic of the FRED pulse is indeed
different from long-lag pulse as shown above.

The anti-correlation between $w$ and photon flux shown in our
analysis is also well established. It is another property of pulse
rather than bursts itself. Whereas Ryde (2005) found an inverse
relation between flux and lag. It is interesting that the three
quantities correlate with each other. If pulse width and spectral
lag can be viewed as mutual surrogates indeed the anti-correlated
relation between $w$ and photon flux must be established. Similar
result given by Hakkila et al. (2008) indicated that there has a
correlation between the pulse duration $w$ and isotropic pulse peak
luminosity. These pulse properties may give a useful constrain on
theoretical model. Therefore, the correlation flux versus $w$ shown
in Figure 7 is established in our sample.

The correlation flux vs. $w$ shown in Figure 7 seems neat, and we
might suspect that it is just the result of a selection bias since
other pulses, such as short-dim pulses are likely under-represented?
Paper I examined the fluence hardness ratios integrated over the
whole burst, with the split at $\tau_{lag} < 1$ s in the range $0.5
< F_{peak} < 2.0$ to see if there is any difference between long-lag
and short-lag dim bursts. They found that the dim short-lag bursts
have slightly harder spectra than the dim long-lag bursts in the
same peak flux range. The analysis of Shahmoradi \& Nemiroff (2010)
showed that simple hardness ratios are good estimator for the
spectral peak energy in GRBs and it is independent of the type of
the burst, whether long-duration GRB or short-duration. So the
short-lag dim bursts should be greater peak energy than that of
long-lag dim bursts. Whereas the mean peak energies of the burst
spectra are correlated with intensity (flux): lower intensity groups
of burst spectra exhibit a lower average peak energy (Mallozzi et
al. 1995). Hence it is suggested that the short-lag dim bursts with
greater peak energy might have higher intensity. Although the above
properties of correlation are bursts rather than pulses we still
think the correlation should be exist among pulses since Borgonovo
\& Bj\"{o¡§}rnsson (2006) showed that the overall properties of a
burst is determined mainly by the properties of pulses. Therefore,
we tend to believe it is not a selection bias even if we can not
give a test with a short-dim pulse sample.

The evident anti-correlation between the evolutionary slope during
the pulse decay phase and pulse asymmetry seems to show pulses with
short rise and very long decay times tend to more slower decay of
$E_{p}$. The tendency appears to suggest the pulses might represent
external shocks capable of initiating afterglow (Hakkila et al.
2008).

\section{Acknowledgments}
We thank the anonymous referee for constructive suggestions. This
work was supported by the Science Fund of the Education Department
of Yunnan Province (08Y0129), the National Natural Science
Foundation of China (No. 10778726), the Natural Science Fund of
Yunnan Province (2009ZC060M).

\clearpage
\end{document}